\documentclass[12pt]{article}

\usepackage{amsmath}
\usepackage{amssymb}
\usepackage[cp866]{inputenc}
\usepackage{epsfig}
\begin{document}

\setcounter{page}{1}

\vspace{.5cm}

\begin{center}
{\LARGE \bf 
The Shapes of Pulsar Radio Pulses in the Plane
of the Sky}
\end{center}

\begin{center}
H.L.~Hakobyan$^1$, V.S.~Beskin$^{1,2*}$\\

\vspace{1cm}

{\it $^{1}$Moscow Institute of Physics and Technology, State University,\\ Institutskii 9, Dolgoprudny, 141700 Russia}\\
{\it $^{2}$Lebedev Physical Institute, Russian Academy of Sciences,\\ 
Leninskii pr. 53, Moscow, 119991 Russia}\\

\vspace{.4cm}


\vspace{.4cm}

{\small \textit
Will be published in Astronomy Reports
\\
 Translated by  D. Gabuzda
}

\end{center}

\vspace{1cm}

{\small
{\bf Abstract} -- The sizes of pulsar radio pulses in the plane of the sky are determined. This is important not only
in relation to the possibility of directly resolving the radio pulses spatially, but also for verifying and placing
constraints on existing theories of the radio emission. The shape of the pulse radio image and its variation
as a function of the pulse phase for the extraordinary (X) and ordinary (O) modes are determined for the
simple case of a dipolar magnetic field. Images are obtained for pulsars with various angles between their
magnetic and rotational axes, and also for various model parameters, such as the height of the emission,
the size of the emission region as a function of the height, and the Lorentz factor of the secondary plasma.
It is shown that the size of the resulting image is appreciably smaller at the center of the pulse than at its
edges.
}
\noindent

\vspace{1cm}
\noindent
\\

\noindent
$^*$ Email: beskin@lpi.ru

\newpage

\begin{center}

{\Large 1. INTRODUCTION}

\end{center}

In spite of the absence of a generally accepted, consistent theory for the coherent emission of radio pulsars, the fully workable hollow-cone model developed over the nearly half a century these objects have been studied is able to explain the main geometrical properties of the mean pulses ~\cite{RC, Smith, MT}. This model is based on the hypothesis that the beam of the radio emission repeats the density profile of the secondary electron-positron plasma, whose creation should be suppressed near the magnetic axis, where the radius of curvature of the magnetic-field lines is appreciably higher than it is far from this axis ~\cite{Sturrock, RS}.

According to current ideas, the secondary plasma is created near the magnetic poles of the neutron star, where there is a strong longitudinal electric field. The primary particles accelerated by this field move along curved trajectories, emitting curvature radiation at the characteristic frequency $\omega_{\rm cur} \sim (c/R_{\rm c})\gamma^3$, which falls in the gamma-ray range ($R_{\rm c}$  is the radius of curvature and $\gamma$ is the Lorentz factor of the particles). The next process in the chain is the conversion of gamma-ray photons into electron-positron pairs, $\gamma+B\rightarrow e^{+}+e^{-}+B$, which becomes possible only in a curved magnetic field; photons moving along the magnetic field are not able to be transformed into electron-positron pairs~\cite{LL}.

We conclude that the birth of secondary particles will be suppressed in a linear magnetic field (more exactly, in a magnetic field with a large radius of curvature), since both the intensity of the curvature radiation and the direction of propagation relative to the magnetic field become small. Therefore, the density of secondary particles should be lower in the central region of a polar cap than at its boundary. The radio emission of pulsars is usually thought to be associated with outflowing plasma.

The structure of the radio-emission beam has been studied in considerable detail, both theoretically and observationally (see, e.g.,~\cite{WJ, HR}). For example, the
dependence of the mean-profile widths on the pulsar period $P$ and the observing frequency $\nu$ has been derived from observations~\cite{R, Gil}. These results showed
that the radio emission should be generated at heights $R_{\rm m}$ of the order of $10$--$100R$, where $R$ is the radius of the neutron star. Much attention has been paid recently to the construction of a consistent theory for the propagation of waves in the magnetosphere~\cite{LP, Lai, BPh} that can explain the main polarization properties of the emission beam.
 
On the other hand, the shape of the radio signal in
the plane of the sky has not been studied previously.
The reason is that, even if the spatial size of the
image is several tens of neutron-star radii (of the order
of $10^7$ cm), the corresponding angular size for the
nearest pulsars (with distances of order $100$ pc) is
only $10^{-4}$ microarcseconds, which cannot be resolved
by current instruments. Therefore, the only means
by which it may be possible to estimate the linear
size of the radiating region is through interstellar
scintillation~\cite{Gwinn, PRAO1, PRAO2}. Even then, it is only possible
to obtain upper limits in most cases. Only recently,
with the launch of the "RadioAstron" space radio
telescope~\cite{UFN_RA} has the possibility of directly resolving
the spatial images of radio pulsars become available.
Therefore, we believe that the time has come to study
this question in more detail.
The main assumptions used to construct the spatial image of the pulse radio emission and the main
parameters influencing the shape of the image are
described in Section 2. Section 3 concerns the methods used to determine the shape of the image and its
dependence on the pulse phase. Finally, Section 4
presents a discussion of the results obtained.
 
\vspace{.4cm}

\begin{center}
{\Large 2. MAIN ASSUMPTIONS}
\end{center}

We will now formulate the main assumptions used
to construct a pulse radio image in the plane of the
sky. Recall that the pulsar radio emission has two
different polarization modes-ordinary (O mode) and
extraordinary (X mode)~\cite{Smith, MT}. The X mode always
propagates along a straight line, while the O mode
deviates from the magnetic axis at small distances
from the neutron star. As was shown in~\cite{BGI}, this
deviation angle depends on the frequency $\nu$ as $\nu^{-0.14}$.
We will consider this in more detail below.

Clearly, the geometrical properties of the observed
radiation will depend on the structure of the emission
beam. For simplicity, we assumed that the radio emission beam repeats the profile of the plasma density at the surface of the polar cap, which we write in
the form

\begin{equation}
\begin{aligned}
n_{\rm e} = \lambda g(r_{\perp})n_{\rm GJ},\\
r_\perp = r \sin\psi_\mathrm{m}.
\end{aligned}
\end{equation}

Here, $\lambda\sim10^3 \div 10^4$ is the multiplicity of the particle
creation,
$n_{\rm GJ} = \mathbf{\Omega B}/2\pi ce$ is the so-called Goldreich-Julian density, and the factor
$g(r_{\perp})$ models the structure of the hollow cone, where $r_{\perp}$ is the
distance from the magnetic axis ($\psi\mathrm{m}$is the angle
between the magnetic axis and the radius vector). For
simplicity, we assumed everywhere that the magnetic
field of the neutron star corresponds to the field of a point dipole (this is clearly valid for distances $R_{\rm m}$, much smaller than the radius of the light cylinder, $R_{\rm L} = c/\Omega$)and did not include the effects of aberration.

As we already noted, there is no generally accepted theory of the radio emission that could be used to determine
$g(r_{\perp})$. Therefore, as in~\cite{BPh}, we used the
one-parameter approximation
\begin{equation}
\label{polarcap}
g(r_{\perp}) = \left[1+\left(\frac{f_0 R_0}{r_{\perp}}\right)^5\right]^{-1}\exp{(-r_{\perp}^2/R_0^2)}.
\end{equation}
Here, $R_0(r) = (\Omega r/c)^{1/2}r$ is the distance from the
magnetic axis to the last open field line, which depends on the radius vector r and determines the opening angle of the beam in the hollow-cone model. The
exponential factor leads to a sharp drop in the radio
intensity when $r_{\perp} > R_0$. The parameter $f_0$ ($0 < f_0 < 1$) specifies the inner radius of the beam cavity, $f_0 R_0$;
i.e., there is no cavity when $f_0 = 0$. Thus, the function
$g(r_{\perp})$ determines the intensity of the emission at a
given emission level $R_{\rm m}$.

Further, since the emission at a given frequency
can be generated over a wide range of heights
$r$, we used the following additional factor in the
parametrization:

\begin{equation}
h(r) = \exp{\left[-A{(r-R_{\rm m})^2\over R_{\rm m}^2}\right]}.
\label{hr}
\end{equation}
$R_{\rm m}$ and $A$ are the second and third parameters of
our problem, and enabled us to take into account
the contributions of different heights. When $A \gg 1$,
the emission is generated only in a narrow range of
heights, $r \approx R_{\rm m}$.

Finally, the fourth key parameter is the width of
the beam for each radiating element relative to the
direction of the magnetic field, $\theta_0$. We took this angle
to be $\theta_0 = 1/\gamma$, where $\gamma$ is the Lorentz factor of the
outflowing plasma. In other words, we assumed that
the emission intensity is proportional to
\begin{equation}
d(\theta)= \exp{\left(-\gamma^2\theta^2\right)},
\end{equation}
where $\theta$ is the angle between the magnetic field and
the direction of the beam propagation. As a result,
the relative contribution of an element along a ray  ${\rm d}l$
to the total intensity at the point ($r, r_{\perp}$) (and for a
ray propagating at an angle $\theta$ to the magnetic field)
is proportional to $g(r_{\perp})h(r)d(\theta)$.

\vspace{.4cm}

\begin{center}
{\Large 3. CONSTRUCTION OF THE IMAGE
IN THE PICTURE PLANE}
\end{center}

\begin{figure}
\centering
\includegraphics[scale=1.3]{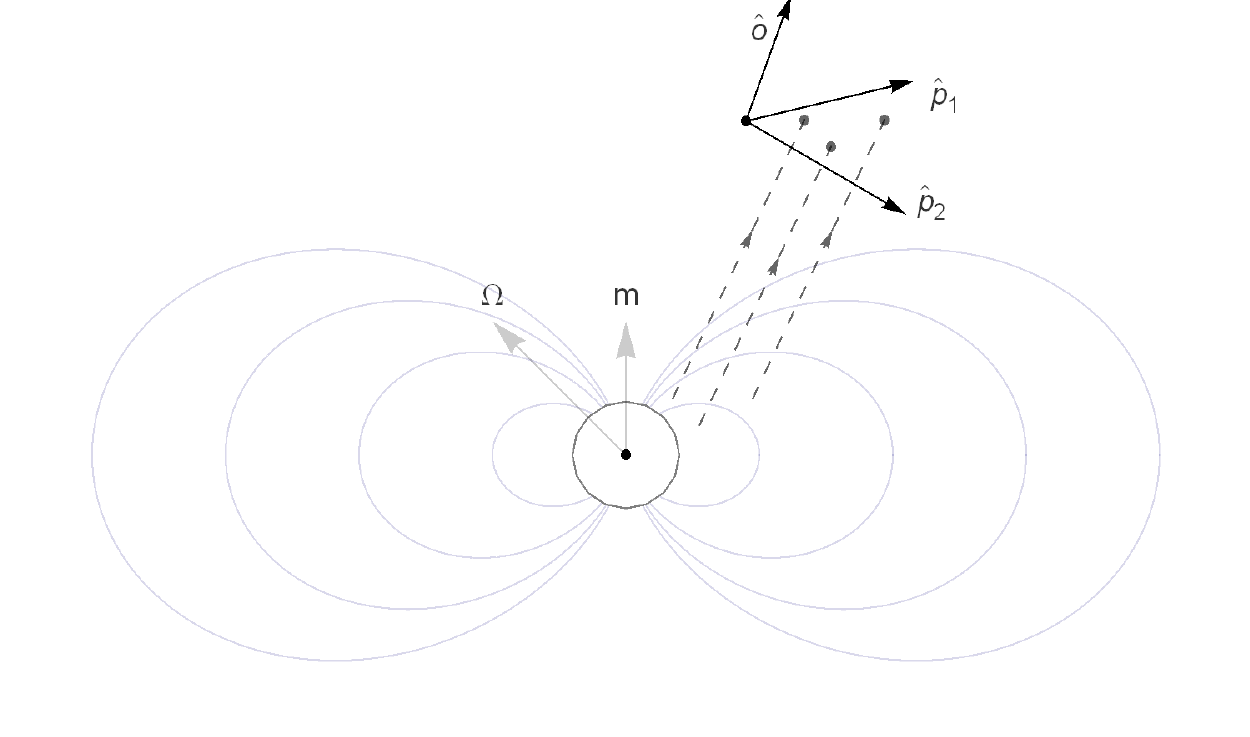}
\caption{Parallel rays that simultaneously intersect the screen (plane of the sky) in the direction toward the observer.}
\label{scheme2}
\end{figure}

Our task is to determine the emission intensity
in the plane of the sky, i.e., for a distant "screen",
for both the X mode and O mode. This reduces to
calculating the integrated intensity along the trajectory of a ray passing through the screen at a point
with coordinates ($a_1$, 
$a_2$) along basis vectors $\mathbf{\hat{p}_1}$ and
$\mathbf{\hat{p}_2}$ perpendicular to the fixed direction toward the observer, $\mathbf{\hat o}$ (Fig.~\ref{scheme2}). In this case, any point in plane
of the sky can be represented

\begin{equation}
\mathbf r(a_1,a_2)= R_{\rm max} \mathbf{\hat o}+a_1\mathbf{\hat{p}_1}+a_2\mathbf{\hat{p}_2},
\label{init}
\end{equation}
where $R_{\rm max}$ is the distance of the screen from the
center of the star (we assumed $R_{\rm max} = 1000R$). The
intensity at a point with coordinates ($a_1, a_2$) can then
be determined by integrating back along the ray trajectory,
\begin{equation}
\label{intens}
I(a_1,a_2) = \int\limits_{\mathbf {r}(t)}^{R_{\rm max}}g(r_{\perp})h(r)d(\theta)\mathrm{d}r
\end{equation}
with fixed $\mathbf{\hat o}$,  $a_1$, and $a_2$ at the surface of the screen.
The image of a radio pulse in the plane of the sky for
a given pulse phase $\phi$ can be obtained by carrying out
the integration (\ref{intens}) along various rays that simultaneously pass through the screen surface in the direction
toward the observer.

Clearly, in spherical coordinates with the $z$  axis
directed along the angular velocity $\mathbf\Omega$, we can write
\begin{equation}
\begin{aligned}
\mathbf{\hat{o}}&=\{\sin{(\chi+\beta)},0,\cos{(\chi+\beta)}\},\\
\mathbf{\hat{p}_1}&=\{\cos{(\chi+\beta)},0,-\sin{(\chi+\beta)}\},\\
\mathbf{\hat{p}_2}&=\{0,1,0\}.
\label{fixvec}
\end{aligned}
\end{equation}

Here and below, $\chi$  is the inclination of the magnetic-
dipole axis $\mathbf m$ to the rotational axis $\mathbf \Omega$, and $\beta$ is the minimum angle between 
$\mathbf{\hat{o}}$ and $\mathbf m$. For convenience, we introduced the dimensionless vectors $\mathbf{\hat{b}}=\mathbf{B}/B$, $\mathbf{\hat{m}}=\mathbf{m}/m$, and  $\mathbf{\hat{n}}=\mathbf{r}/r$. In this case, the vectors that depend on the time $t$ have the form
\begin{equation}
\begin{aligned}
\mathbf{{m}}(t)&= |\mathbf{{m}}|\{\sin{\chi}\cos{(\phi+\Omega(t-t_0))},\sin{\chi}\sin{(\phi+\Omega(t-t_0))},\cos{\chi}\},\\
\mathbf{B}(t )&=-{\mathbf{m}\over r^3}+{3(\mathbf{m},\mathbf{\hat{n}})\mathbf{\hat{n}}\over r^3}.
\label{dynvec}
\end{aligned}
\end{equation}

Naturally, determining the integral (\ref{intens}) requires
knowledge of the ray trajectory $\mathbf{r}(t)$. In a geometrical-
optics approximation, we obtain
\begin{equation}
\begin{aligned}
\frac{\mathrm{d}\mathbf{r}}{\mathrm{d}t}&=\frac{\partial\omega}{\partial\mathbf{k}}
=c \, \frac{\partial}{\partial\mathbf{k}}\left({k\over n}\right),\\
\frac{\mathrm{d}\mathbf{k}}{\mathrm{d}t}&=-\frac{\partial\omega}{\partial\mathbf{r}}
=-c \, \frac{\partial}{\partial\mathbf{r}}\left({k\over n}\right).
\label{geomoptics}
\end{aligned}
\end{equation}

For the linearly propagating X mode ($n\equiv 1$), we have
simply
\begin{equation}
\begin{aligned}
\frac{\mathrm{d}\mathbf{r}}{\mathrm{d}t}&=c \, \frac{\mathbf{k}}{k},\\
\frac{\mathrm{d}\mathbf{k}}{\mathrm{d}t}&=0,
\end{aligned}
\end{equation}
and, based on the initial condition (\ref{init}), we can specify
the trajectory as follows:
\begin{equation}
\begin{aligned}
\mathbf{r}(t)&=(R_{\rm max} - ct)\mathbf{\hat{o}}+a_1\mathbf{\hat{p}_1}+a_2\mathbf{\hat{p}_2},\\
\mathbf{k}&= \mathbf{\hat{o}}= {\rm const}.
\end{aligned}
\end{equation}

The equations of motion are more complex in the case
of the O mode, since the index of refraction $n$ depends
on the angle $\theta$ between the magnetic field and the
wave vector $\mathbf{k}$ in a non-trivial way~\cite{BGI}:
\begin{equation}
n = 1 + {\theta^2\over4}-\left({\theta^4\over16}+\theta_*^4\right)^{1/2}.
\end{equation}
Here,
\begin{equation}
\theta_*^4 = \frac{\omega_\mathrm{p}^2}{\omega^2\gamma^3},
\end{equation}
where the plasma frequency
\begin{equation}
\omega_\mathrm{p}^2 = \frac{4\pi e^2 n_\mathrm{e}}{m_\mathrm{e}},
\end{equation}
will depend on the coordinate $r$ through the particle
number density
\begin{equation}
n_\mathrm{e} = \frac{\lambda}{2\pi c e}g(r_\perp)\left(\mathbf{\Omega}\mathbf{B}\right).
\end{equation}

\begin{figure}[htb]
\centering
\includegraphics[scale=1.2]{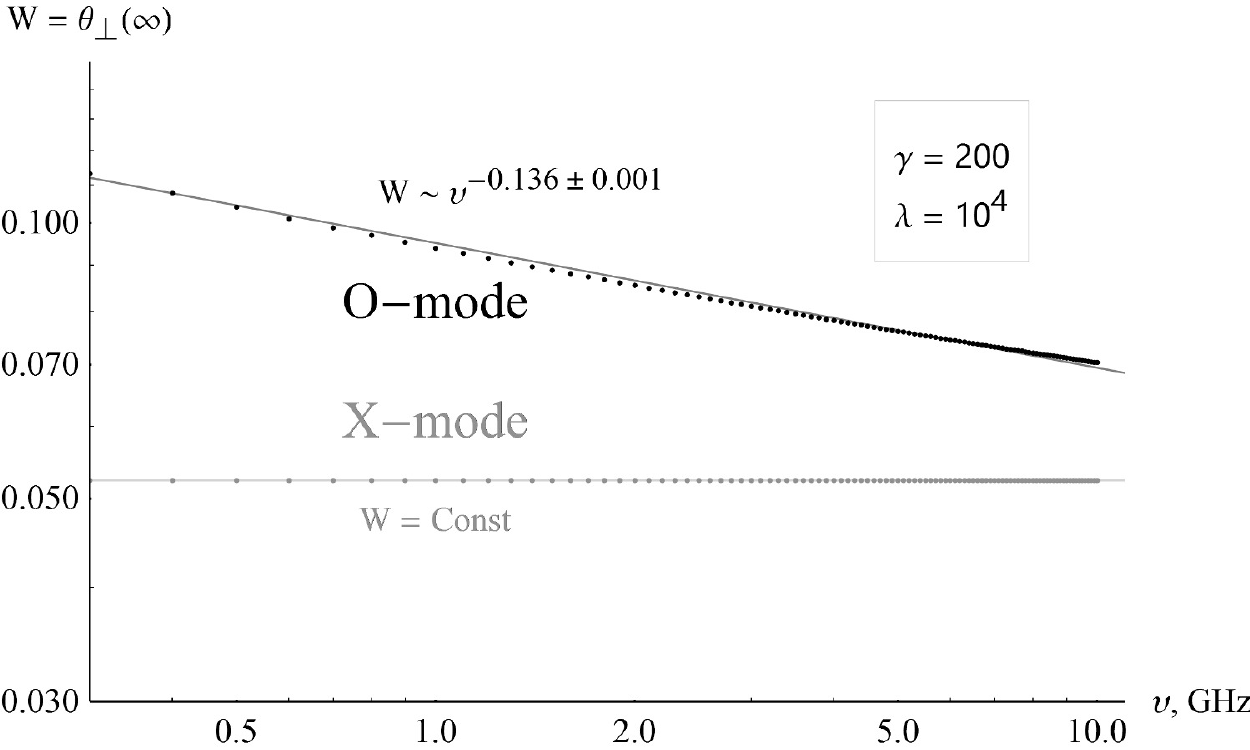}
\caption{Beam width $W$ for the O and X modes for emission from a height of $R_{\rm m} = 10\,R$ for various frequencies $\nu$.}
\label{width}
\end{figure}

As a test, Fig.~\ref{width} compares the numerical simulations for the dependence of the beam width $W$ on the
frequency $\nu$ for the case $g(r_{\perp}) = 1$, $R_{\rm m} = 10\,R$ and
the analytical results obtained in~\cite{BGI} ($W =$ const for
the X mode and $W \propto \nu^{-0.14}$ for the O mode). These
are in very good agreement.

\vspace{.4cm}

\begin{center}
{\Large 4. DISCUSSION}
\end{center}

\begin{figure}[htb]
\centering
\vspace{20pt}
\includegraphics[scale=0.35]{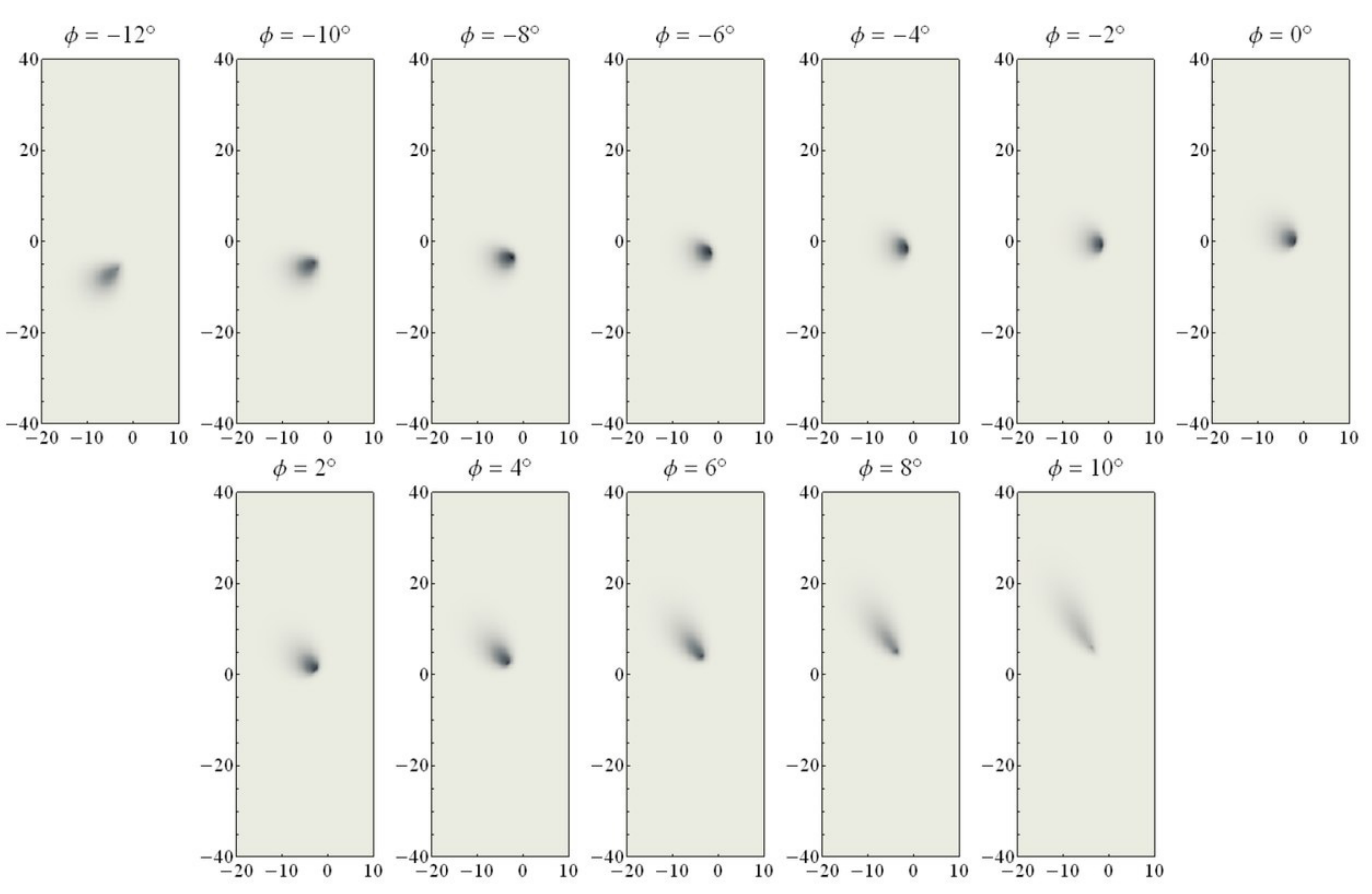}
\caption{
Image of a radio pulse in the plane of the sky for $A=0.1$, $R_m=100$, $\chi=45^\circ$, and $\gamma=100$. The sizes are shown in
units of the neutron-star radius $R$.}
\label{gamma100}
\end{figure}

As an example, Fig.~\ref{gamma100} presents the motion of the
image in the plane of the sky for the O mode from
$\phi=-12^\circ$ to $\phi=10^\circ$ for the case $\gamma=100$. The linear
dimensions are presented in units of the neutron-star
radius $R$. Here and below, we have assumed $\beta=5^\circ$, $\lambda=10^4$, 
$\omega=100~\textrm{MHz}$, $f_0=0.5$ and $P=1~\textrm{s}$. The
emission is slightly supressed at the pulse phase $\phi \approx 0^\circ$ due to the presence of the cavity $f_0$, however, this
suppression is not stronger because the passage is
not central and $\beta\ne0^\circ$. On the other hand, we can
clearly see that the image is strongly elongated at the
edges of the mean profile.
This is easy to understand. Consider the case
when the magnetic axis is perpendicular to the rotational axis ($\chi = 90^{\circ}$), and the observer is located in
the equatorial plane ($\beta = 0^{\circ}$). Since the emission occurs along magnetic-field lines, it can be received only
from points that also lie in the equatorial plane. All the
other field lines will deviate from the equatorial plane
toward the North or the South, so that the emission
from those points will not be detected. Consequently,
in the limiting case $\theta_0 = 1/\gamma \rightarrow 0$, the image in the
plane of the sky will form a line lying in the equatorial
plane, whose length depends on the emission level
$R_{\rm m}$. It is obvious that this elongation will increase
with increasing $\gamma$.

For convenience of analysis, we present plots of
various parameters obtained in our computations.
The most important of these is the mean profile the dependence of the integrated intensity $I$ on the
phase $\phi$. Another important characteristic is the dependence of the linear size of the image in the plane of
the sky on the phase $\phi$. We also present the trajectory
of the intensity peak in the coordinates $a_1$ and $a_2$.
Unless otherwise stated, $A=0.1$, $\chi=45^\circ$, $\gamma=50$, and $R_{\rm m}=50$.

\begin{figure}[htb]
\centering
\includegraphics[scale=0.45]{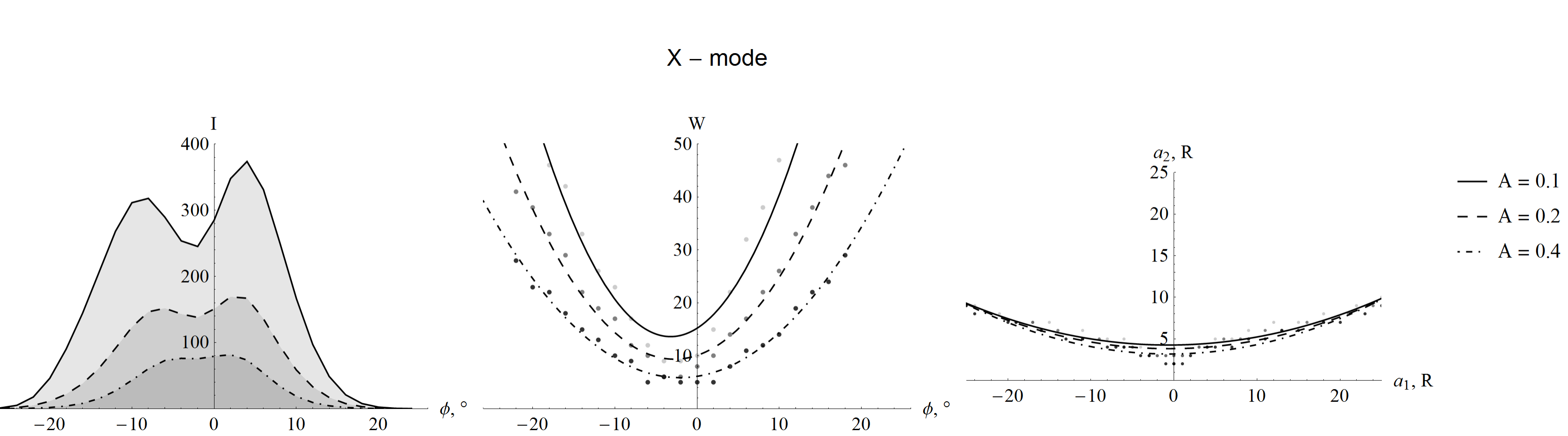}
\caption{Mean profile, image width, and motion of the image center for various values of $A$ for the X mode.}
\label{plotA1}
\includegraphics[scale=0.45]{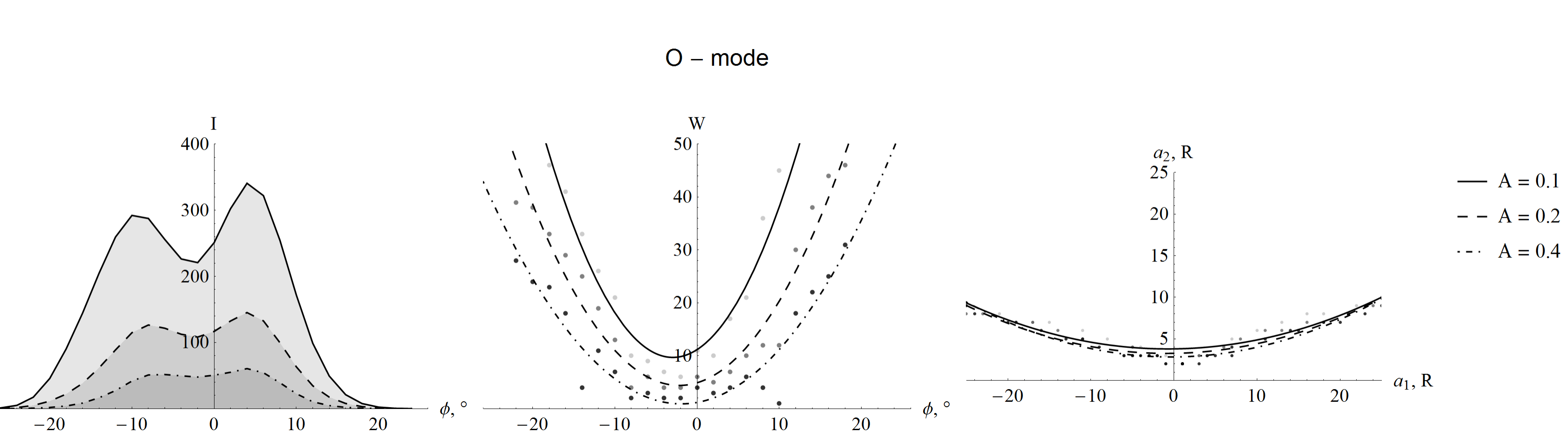}
\caption{Mean profile, image width, and motion of the image center for various values of $A$ for the O mode.}
\label{plotA2}
\end{figure}

\begin{figure}[htb]
\centering
\includegraphics[scale=0.45]{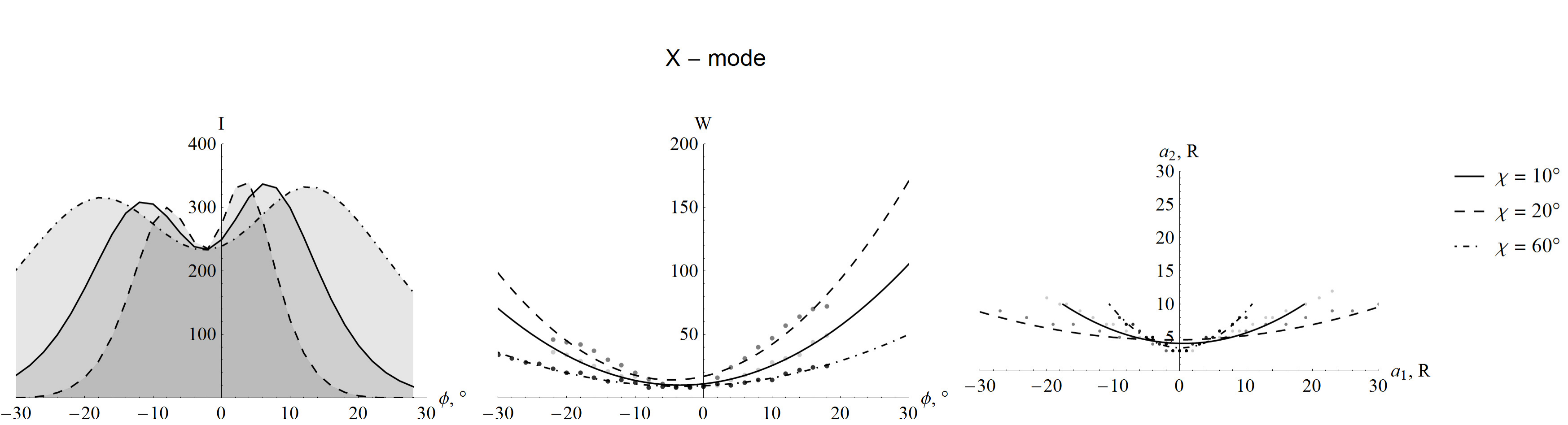}
\caption{Same as Fig.~\ref{plotA1} for various axis inclinations $\chi$.}
\label{plotchi1}
\includegraphics[scale=0.45]{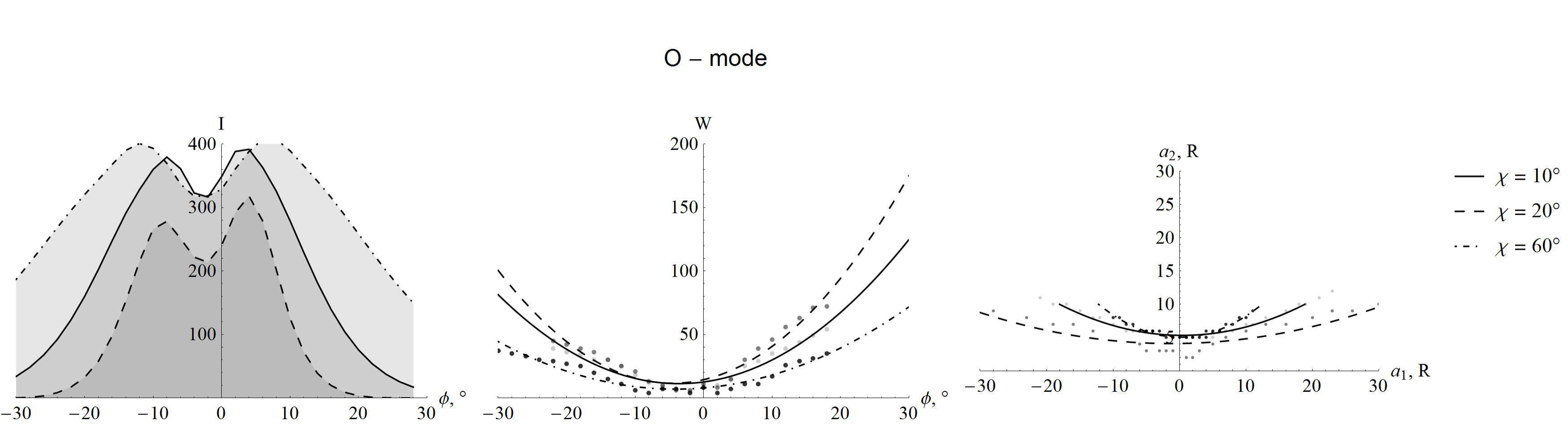}
\caption{Same as Fig.~\ref{plotA2} for various axis inclinations $\chi$.}
\label{plotchi2}
\end{figure}

Figures~\ref{plotA1} and \ref{plotA2} show the mean profile, image
width, and motion of the image center for various
values of $A$, for the O and X modes, respectively. The
parameter $A$ influences the width of the Gaussian (\ref{hr}),
which specifies the dependence of the emission intensity on the height $r$, $h(r)$. As $A$ is increased,
this function becomes more degenerate, so that the
emission emerges from a narrow layer at height $R_{\rm m}$.
This leads to a low integrated intensity and a decrease
in the profile width. In addition, the dependence of
the image width on the phase $\phi$ becomes weaker as
$A$ is increased, since the image size is larger the
more intense the emission at low heights. At the
same time, the motion of the image center is virtually
independent of $A$.

\begin{figure}[htb]
\centering
\includegraphics[scale=0.45]{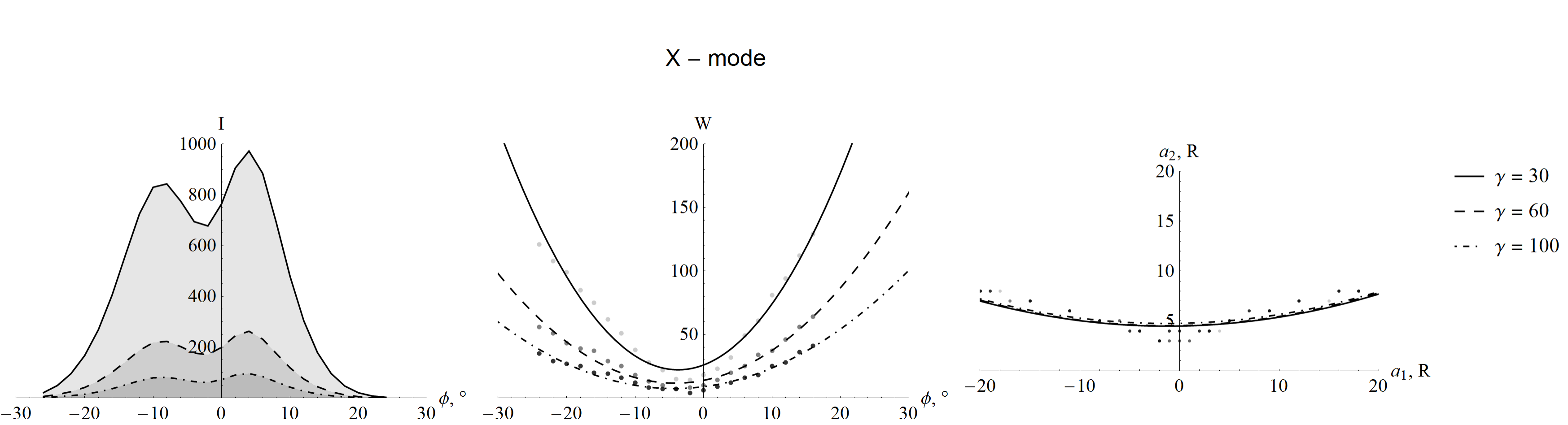}
\caption{Same as Fig.~\ref{plotA1} for various values of $\gamma$.}
\label{plotgamma1}
\includegraphics[scale=0.45]{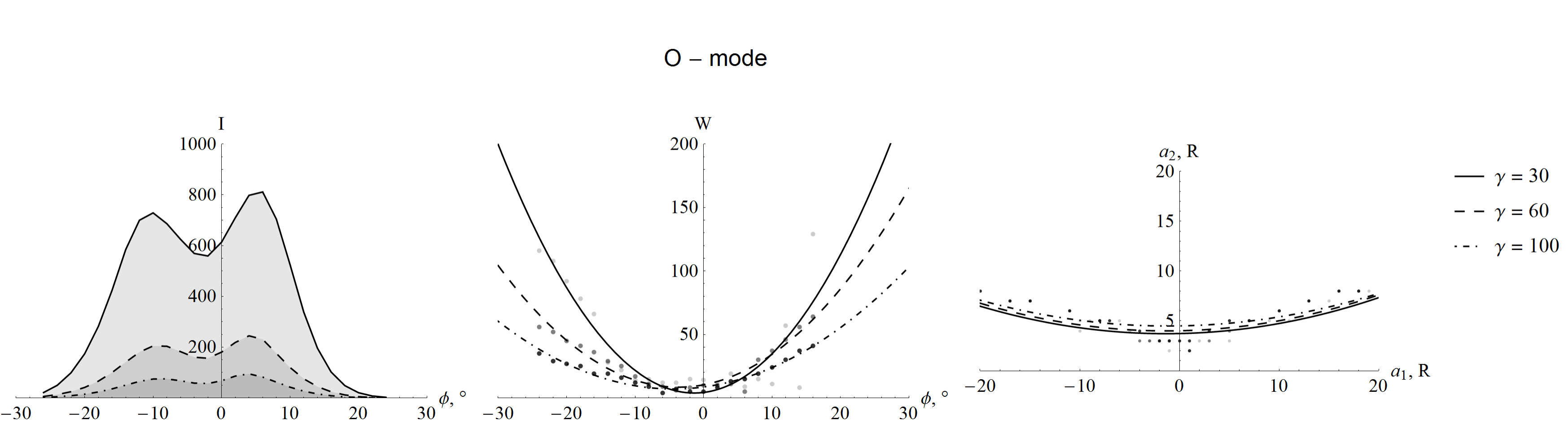}
\caption{Same as Fig.~\ref{plotA2} for various values of $\gamma$.}
\label{plotgamma2}
\end{figure}

Figures~\ref{plotchi1} and \ref{plotchi2} show the mean profile, image
width, and image-center motion for various axis inclinations $\chi$. As expected, this geometric factor strongly
influences the width of the mean profile. When $\chi$ is
small, the mean profile propagates over an appreciable part of the full period. On the other hand, when
$\chi$ is close to $90^\circ$, the dependence of the image size
on the phase $\phi$ strengthens, as is also obvious from
geometrical arguments (the projection of the polar
cap varies more strongly than for $\chi=10^\circ$). Finally,
the motion of the image center depends strongly on
$\chi$. In the case of small inclinations, this trajectory
degenerates into an elongated ellipse (the pulse is
visible over the entire period), while it forms a line for
inclinations close to $90^\circ$.

\begin{figure}[tb]
\centering
\includegraphics[scale=0.45]{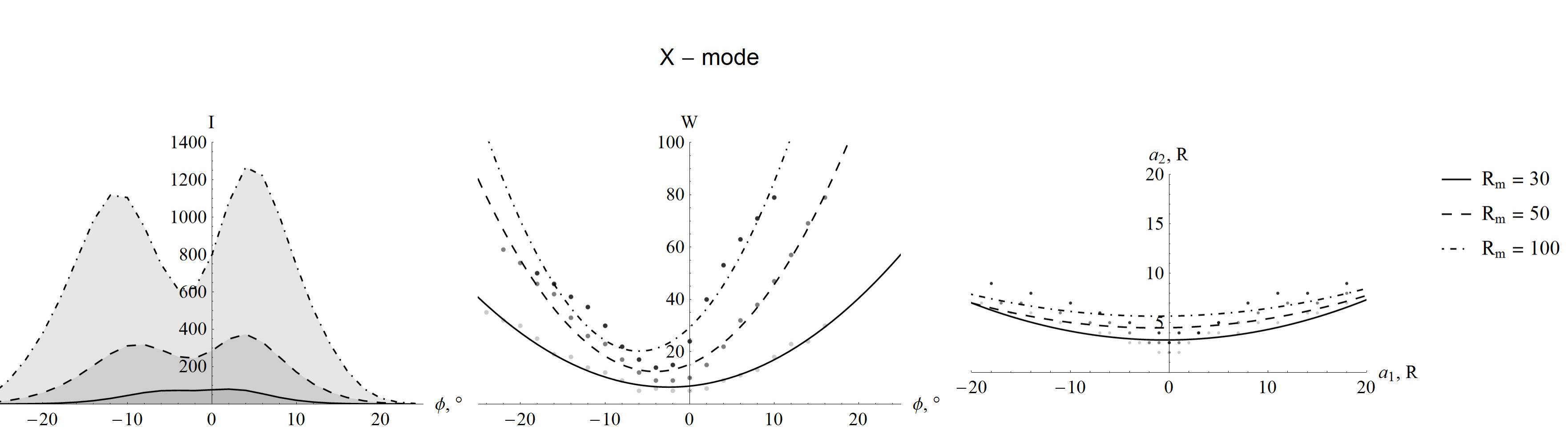}
\caption{Same as Fig.~\ref{plotA1} for various values of $R_{\rm m}$.}
\label{plotRm1}
\includegraphics[scale=0.45]{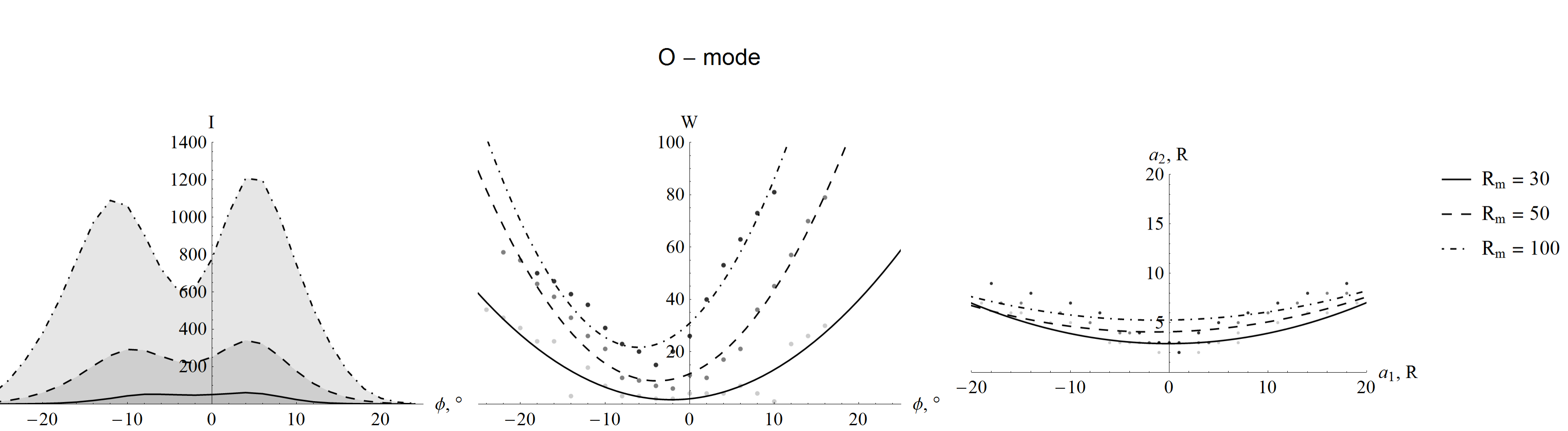}
\caption{Same as Fig.~\ref{plotA2} for various values of $R_{\rm m}$.}
\label{plotRm2}
\end{figure}

Figures~\ref{plotgamma1} and \ref{plotgamma2} present the mean profile, image width, and image-center motion for various values of
$\gamma$. In the profile and image-width plots, increasing
$\gamma$ has the same effect as increasing $A$: a decrease
in the effective angle at which the radio emission is
generated, causing the total intensity to fall. At the
same time, as for $A$, the motion of the image center
depends only weakly on $\gamma$.

Finally, Figures~\ref{plotRm1} and \ref{plotRm2} show the mean profile,
image width, and image-center motion for various
values of $R_{\rm m}$. In the case of large $R_{\rm m}$ and $A=0.1$, a
wide band about $R_{\rm m}$ radiates, extending downward
to $R$ and upward to $10R_{\rm m}$. On the contrary, in the
case of small $R_{\rm m}$, the radiation at high altitudes is
suppressed. The pulsar radiates over a much wider
interval of phase in the case of large $R_{\rm m}$. This gives
rise to a sharp dependence of the image size on the
height when $R_{\rm m}\sim100\,R$. Again, the motion of the
image center depends only weakly on $R_{\rm m}$.

Thus, even for fairly high emission heights $R_{\rm m}\sim100\,R$, the size of the radio-pulse image does not
exceed several neutron-star radii, of order $10^7$ cm.
It is therefore not surprising that it has not been
possible to resolve pulsar images. However, our main
conclusion that the size of the image at the center
of the pulse should be appreciably smaller than at
the pulse edges is supported by recent data on interstellar scintillation~\cite{Gwinnn}. Finally, Pen et al.~\cite{velocity}
have recently reported direct measurements of the
velocity of a pulsar image on the plane of the sky of
order 1000 km/s -- close to the velocity with which the
image moves in Fig.~\ref{gamma100}.

\vspace{.4cm}

\begin{center}
{\Large ACKNOWLEDGEMENT}
\end{center}

The authors thank Ya.N. Istomin, Yu.Yu. Kovalev,
M.V. Popov, and C.R. Gwinn for useful discussions.
This work was supported by the Russian Foundation
for Basic Research (project 14-02-00831).

\end{document}